\documentclass{PoS}
\usepackage{epsfig}

\title{Extraction of Compton Form Factors from DVCS data}

\ShortTitle{Extraction of Compton Form Factors from DVCS data}

\author{\speaker{Michel Guidal}\\
        CNRS/IPN Orsay\\
        E-mail: \email{guidal@ipno.in2p3.fr}}

\abstract{
We present the results of a fitter code which aims at extracting 
Compton Form Factors (CFFs) from DVCS (Deep Virtual Compton Scattering) 
experimental data, in a largely model-independent way. CFFs are 
linked to GPDs (Generalized parton Distributions) and are the quantities
which are directly measurable. The data that we
have analyzed are from JLab and HERMES experiments. We obtain some first 
important constraints on the $H$ and $\tilde{H}$ CFFs. The kinematical 
dependences ($x_B$, $t$) of these CFFs provide some new insights on nucleon 
structure.}

\FullConference{35th International Conference of High Energy Physics\\
		 July 22-28, 2010\\
		 Paris, France}

\begin{document}

Generalized Parton Distributions allow us to describe 
the structure of the nucleon in a very rich and unprecedented way. 
Among other things, they contain the correlations between the (transverse) 
position and (longitudinal) momentum distributions of the partons in the nucleon, 
they allow us to derive the orbital momentum contribution of partons to the nucleon's 
spin and they provide an access to the nucleon's ($q\bar{q}$) content. 
Experimentally, GPDs are the most simply accessed  through the exclusive 
leptoproduction of a photon (DVCS~: $ep\to e'p'\gamma$) or of 
a meson. We refer the reader to Refs.~\cite{muller,ji,rady,goeke,
revdiehl,revrady,myppnp} for the original theoretical articles and 
recent comprehensive reviews on GPDs and for details on the theoretical formalism.

At QCD leading twist and leading order approximation, there are four independent 
nucleon GPDs which can be accessed in the DVCS process: $H$, $E$, $\tilde{H}$ and 
$\tilde{E}$. These four GPDs depend on three variables $x$, $\xi$ and $t$, of
which only two are experimentally accessible: $t$ and $\xi$, where 
$\xi$ is related to the standard Deep Inelastic Bjorken variable
$x$$_B$ through the formula: $\xi=\frac{x_B}{2-x_B}$.
This is why only CFFs, which are weighted integrals of GPDs over $x$ 
or combinations of GPDs at the line $x=\xi$, can in general be 
extracted from DVCS experiments. In our notation which was introduced 
and used in Refs.~\cite{fitmick,fithermes,fittsa,fitall}, there are eight 
CFFs which are:
\begin{eqnarray}
&&H_{Re}=P \int_0^1 d x \left[ H(x, \xi, t) - H(-x, \xi, t) \right] C^+(x, \xi),\label{eq:eighta} 
\\
&&E_{Re}=P \int_0^1 d x \left[ E(x, \xi, t) - E(-x, \xi, t) \right] C^+(x, \xi),\label{eq:eightb} 
\\
&&\tilde{H}_{Re}=P \int_0^1 d x \left[ \tilde H(x, \xi, t) + \tilde H(-x, \xi, t) \right] C^-(x,
\xi),\label{eq:eightc} 
\\
&&\tilde{E}_{Re}=P \int_0^1 d x \left[ \tilde E(x, \xi, t) + \tilde E(-x, \xi, t) \right] C^-(x,
\xi),\label{eq:eightd} 
\\
&& H_{Im}=H(\xi , \xi, t) - H(- \xi, \xi, t),\label{eq:eighte} \\
&& E_{Im}=E(\xi , \xi, t) - E(- \xi, \xi, t),\label{eq:eightf} \\
&& \tilde{H}_{Im}=\tilde H(\xi , \xi, t) + \tilde H(- \xi, \xi, t) 
\\
&&\tilde{E}_{Im}=\tilde E(\xi , \xi, t) + \tilde E(- \xi, \xi, t)\label{eq:eighth} 
\end{eqnarray}
with 
\begin{equation}
C^\pm(x, \xi) = \frac{1}{x - \xi} \pm \frac{1}{x + \xi}.
\end{equation}

In Refs.~\cite{fitmick,fithermes,fittsa,fitall}, we have developed a largely
model-independent fitting procedure which, at a given experimental
($\xi$, $-t$) kinematic point, takes the CFFs as free parameters and
extracts them from DVCS experimental observables using the well
established DVCS theoretical amplitude~\cite{kirch,vgg1}. 
This task is not trivial. Firstly, one has to fit seven~\footnote{Guided by
theory considerations, we actually neglect $\tilde{E}_{Im}$ in our work.} parameters 
from a limited set of data and observables, which leads in general to an
under-constrained problem.  However, as some observables are in
general dominated by a few particular CFFs, one can extract 
a few specific CFFs. Secondly, in addition to the particular DVCS process 
of direct interest, there is another mechanism which 
contributes to the $ep\to e'p'\gamma$ process and whose amplitude interferes 
with the DVCS amplitude. This is the Bethe-Heitler (BH) 
process where the final state photon is radiated by the incoming or scattered 
electron and not by the nucleon itself. However, it is  
precisely known and calculable given the nucleon form factors.

With our fitting algorithm, we have managed to determine in previous works~:
\begin{itemize}
\item the $H_\mathrm{Im}$ and $H_\mathrm{Re}$ CFFs. at
  $<x_B>\approx0.36$, and for several $t$ values, by
  fitting~\cite{fitmick} the JLab Hall A proton DVCS beam-polarized
  and unpolarized cross sections~\cite{franck},
\item the $H_\mathrm{Im}$ and $\tilde{H}_\mathrm{Im}$ CFFs, at
  $<x_B>\approx 0.35$ and $<x_B>\approx 0.25$, and for several $t$
  values, by fitting~\cite{fittsa} the JLab CLAS proton DVCS
  beam-polarized and longitudinally polarized target spin
  asymmetries~\cite{fx,chen},
\item the $H_\mathrm{Im}$, $H_\mathrm{Re}$ and $\tilde{H}_\mathrm{Im}$ CFFs, 
  at $<x_B>\approx 0.09$, and for several $t$ values, by fitting~\cite{fithermes,
  fitall} a series of HERMES beam-charge, beam-polarized, transversely and 
  longitudinally polarized target spin asymmetry moments~\cite{ave,hermes,hermesaul,
  davidthesis}.
\end{itemize}

\begin{figure}[htb]
\epsfxsize=12.cm
\epsfysize=12.cm
\epsffile{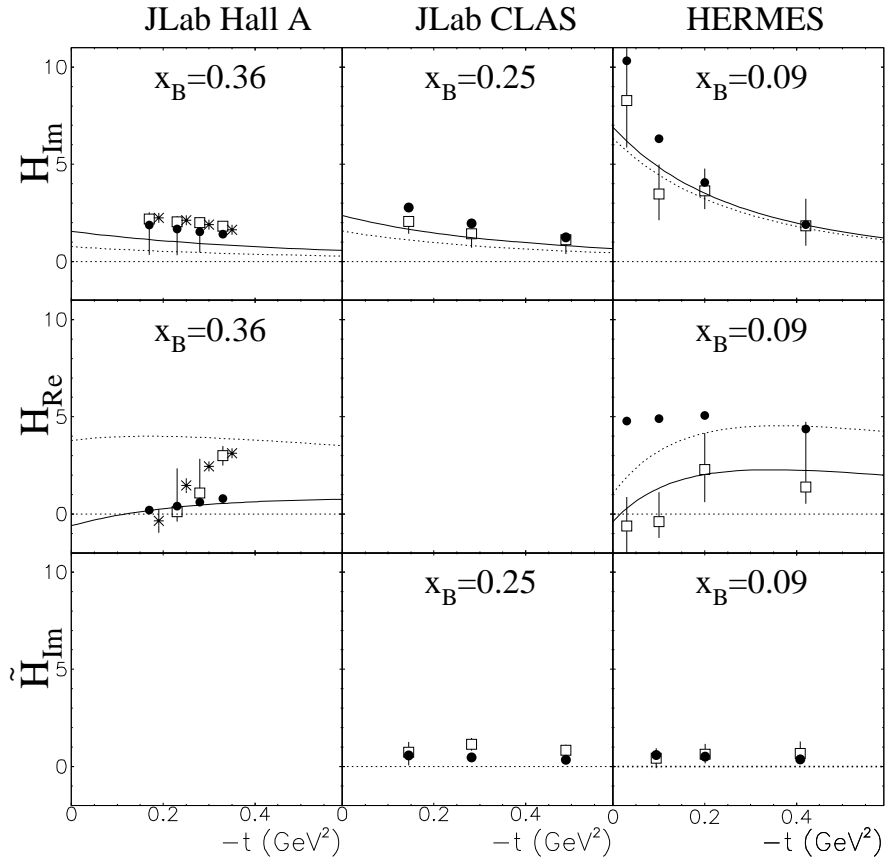}
\caption{The $H_\mathrm{Im}$, $H_\mathrm{Re}$ and $\tilde{H}_\mathrm{Im}$
CFFs, as defined in Eqs.1, 5 and 7, as a function of $-t$. The empty squares 
show the results of our works, the stars the result of the CFF fit of 
Ref.~\cite{herve}, the curves the results of the model-based fit of 
Ref.~\cite{fitmuller} and the solid points show the predictions of the 
VGG model~\cite{vgg1,goeke,gprv}.}
\label{fig:allfit}
\end{figure}


In Fig.~\ref{fig:allfit}, we compile all our results, each panel having 
the same scales for ease of comparison. The empty squares show the results of our works, the stars the result of the CFF fit of Ref.~\cite{herve},
the curves the result of the model-based fit of Ref.~\cite{fitmuller}
(solid: without the Hall A data of Ref.~\cite{franck} and dashed: including 
the Hall A data in the fit) and the solid points show the predictions of the 
VGG model~\cite{goeke,vgg1,gprv}.
Except for $H_\mathrm{Re}$ where there are marked differences
between the different approaches, it seems that all these works show the same
trends within error bars.

Our results have average uncertainties of the order of 30\%. 
This is due to the limited precision of the data and/or the limited number of 
experimental observables to be fitted. Obviously, having more observables to 
fit simultaneously and more precise data (which can be foreseen in the near future) can only reduce these uncertainties. Also, one has to keep in mind 
that we keep in our fits all seven CFFs as free parameters~\footnote{To be 
precise, in our work, the CFFs are actually bound to vary in a space 
$\pm 5$ times some reference model values which should be a priori
a very conservative hypothesis.}. If, guided by some theoretical 
considerations, one can remove some of the CFFs 
from the fit and thus reduce the number of free parameters, error bars on 
the results will obviously diminish. For instance, in Ref.~\cite{herve}
(stars on Fig.~\ref{fig:allfit}), all GPDs but $H$ have been neglected
resulting in smaller uncertainties (an additional error has then
to be introduced in order to take into account the neglect of the
other GPDs; an attempt of the estimation of such addditional error has 
been done in Ref.~\cite{herve}). For the present time, in our
approach, our uncertainties reflect all our ignorance on all GPDs other 
than the ones which come out from our fits and their full potential 
influence. In particular, we found in Ref.~\cite{fittsa} that the central
value on the fitted $H_\mathrm{Im}$ CFF would vary by a factor of $\approx$ 3
whether one would fit the JLab Hall A proton DVCS cross sections~\cite{franck}
by taking into account only the $H$ GPD or by taking into account all four GPDs. 
  
In Fig.~\ref{fig:allfit}, some general features and trends can be 
distinguished. We comment them briefly in the following:

\begin{itemize}
\item Concerning $H_\mathrm{Im}$, it seems that, at fixed $-t$, 
this CFF increases as $x_B$ decreases (i.e. going from JLab to HERMES kinematics). This is reminiscent of the $x$-dependence of the standard 
proton unpolarized parton distribution as measured in DIS, to which 
$H$$_\mathrm{Im}$ reduces in forward kinematics ($\xi=t=0$).
Another feature is that the $t-$slope of $H$$_\mathrm{Im}$ seems to increase 
with $x_B$ decreasing. This could then suggest that low-$x$ quarks (the ``sea") 
would extend to the periphery of the nucleon while the high-$x$ (the 
``valence") would tend to remain in the center of the nucleon. Indeed, 
the $t$-dependence of GPDs can be interpreted as a reflection of the spatial 
distribution of some charge in some specific 
frame~\cite{Burkardt:2000,Diehl:2002he,pire}. 

\item $H_\mathrm{Re}$ has a very different $t$-dependence than $H_\mathrm{Im}$
both at JLab and at HERMES energies: while $H$$_\mathrm{Im}$ decreases with $-t$ 
increasing, $H_\mathrm{Re}$ increases (at least up to $-t\approx$ 0.3 
GeV$^2$) and may change its sign, starting negative 
at small $-t$ and reaching positive values at larger $-t$: all four 
aproaches (empty squares, stars,
solid points and solid curves) show this ``zero-crossing" at JLab kinematics 
while only our CFF fitting work tends to show it for HERMES kinematics.
We also notice that both VGG and the dashed curve of the model-based fit 
of Ref.~\cite{fitmuller} overestimate our fitted values for $H_\mathrm{Re}$.

\item Concerning $\tilde{H}_\mathrm{Im}$, we notice that it is in general
smaller than $H_\mathrm{Im}$, which can be expected for a polarized
quantity compared to an unpolarized one. There is very little $x_B$ dependence. 
The $t$-dependence is also rather flat. The weaker $t$-dependence 
of $\tilde{H}_\mathrm{Im}$ compared to ${H}_\mathrm{Im}$ suggests
that the axial charge (to which the $\tilde{H}$ GPD is related) has a narrower 
distribution in the nucleon than the electromagnetic charge.
\end{itemize}
 
To summarize, in this short report, we have presented 
the results of our fitting code, based on the leading twist 
and leading order QCD DVCS handbag diagram amplitude (and BH), 
which aims at extracting CFFs from DVCS data (in the quark sector).
We have extracted in a largely model-independent way some numerical 
constraints on three CFFs~: 
$H$$_\mathrm{Im}$, $H_\mathrm{Re}$ and $\tilde{H}_\mathrm{Im}$, 
which hint at some original features of the nucleon structure.
Where applicable, we have compared our results with other approaches 
which in general show the same trends as the ones we found.

We are very thankful to V. Burkert, K. Kumericki, H. Moutarde and 
D. M\"uller for useful discussions.

\end{document}